\documentclass[11pt]{aastex}

\input{epsf}

\newcommand{\gsim}{\mathrel{\hbox{\rlap{\hbox{\lower4pt\hbox{$\sim$}}}\hbox{$>$}}}}
\newcommand{\lsim}{\mathrel{\hbox{\rlap{\hbox{\lower4pt\hbox{$\sim$}}}\hbox{$<$}}}}

\shortauthors{Hoyle \& Vogeley} 
\shorttitle{Voids in the PSCz and UZC}

\begin{document}

\title{Voids in the PSCz Survey and the Updated Zwicky Catalog }

\author{Fiona Hoyle \& Michael S. Vogeley \\ 
Department of Physics, Drexel University, 
3141 Chestnut Street, Philadelphia, PA 19104 \\ 
\email{hoyle@venus.physics.drexel.edu, vogeley@drexel.edu}
}

\begin{abstract}
We describe an algorithm to detect voids in galaxy redshift
surveys. The method is based on the {\tt void finder} algorithm of
El-Ad \& Piran.
We apply a series of tests to determine how accurately we are able to
recover the volumes of voids using our detection method. We simulate
voids of different ellipticity and find that if voids are
approximately spherical, our algorithm will recover 100\% of the
volume of the void. The more elliptical the void, the smaller the
fraction of the volume we can recover. We 
insist that voids lie completely within the survey. Voids close
to the edge of the survey will therefore be underestimated in volume. 
By considering a deeper sample, we estimate the maximal sphere diameters are 
correct to within 30\%.

We apply the algorithm to the Point Source Catalogue Survey (PSCz) and
the Updated Zwicky Catalog (UZC). The PSCz survey is an almost all-sky 
survey with objects selected from the IRAS catalog.
The UZC covers a smaller area of sky but is optically selected and samples
the structures more densely. 
We detect 35 voids in the PSCz and 19 voids in the UZC with
diameter larger than 20$h^{-1}$Mpc. Using this minimum size threshold,
voids have an average effective diameter 
of 29.8$\pm 3.5 h^{-1}$Mpc (PSCz) and 29.2$\pm 2.7 h^{-1}$Mpc (UZC) and that 
they are underdense
regions with $\delta \rho / \rho$ values of -0.92$\pm0.03$ (PSCz) 
and -0.96$\pm0.01$ (UZC) 
respectively. Using this quite stringent threshold for void definition, 
voids fill up to 40\% of the volume of the universe.

\end{abstract}

\keywords {cosmology: large-scale structure of the universe --
cosmology: observations -- galaxies: distances and redshifts --
methods: statistical}

\section{Introduction}

The distribution of galaxies in redshift surveys
reveals vast regions of space 
that seem to be avoided by galaxies. These are termed voids, although voids
may not be completely empty, but may harbor a few isolated galaxies.
During the 1970's, the use of galaxy redshift surveys to trace the
large scale structure began in earnest. Gregory \& Thompson (1978) 
showed evidence for the existence of superclusters and void regions with radii 
larger than 20$h^{-1}$Mpc \footnote{we adopt the
convention that $H_{\circ}=100 h $km s$^{-1}$ Mpc$^{-1}$}
using pencil beam surveys 
directed toward the Coma and Perseus cluster. 
Einasto, Joeveer \& Saar (1980) discuss the chain-like distribution of 
galaxies and galaxy clusters and the existence of empty cell structures.
Kirshner et al. (1981)
discovered a void in Bo{\"o}tes that is 50$h^{-1}$Mpc  in diameter.
The ubiquity of voids and their importance in the large scale distribution of
galaxies was first clearly shown by results from the
Center for Astrophysics surveys (Davis et al. 1992; de Lapparent, Geller, \& 
Huchra 1986; Geller \& Huchra 1989) and the Southern Sky Redshift survey
(da Costa et al. 1988, 1994; Maurogordato et al. 1992).
Subsequent larger surveys at a variety of wavelengths have
confirmed these results and have shown that no larger voids are seen at 
similar density contrast.
See the review article
by Rood (1998) and references therein for a discussion of the
history of void detection and interpretation.

Voids appear to have little or no structure within
them. Peebles (2001) discusses 
an apparent discrepancy between the cold dark matter
model and observations.
The cold dark matter model predicts that there 
should be matter and,
hence galaxies, primarily dwarfs, within voids
(Dekel \& Silk 1986; Hoffman, Silk \& Wyse 1992). However, studies of
different types of galaxies show that they all seem to trace the same
structures. Grogin \& Geller (1999, 2000) identify a sample of void galaxies in the CfA2 survey and find evidence for environmental dependence of galaxy
properties, although most of these galaxies in this sample avoid the
centers of the voids.
Surveys of dwarf galaxies indicate that they trace the
same overall structures as `normal' galaxies (Bingelli 1989) and
pointed observations toward void regions also fail to detect a
significant population of faint galaxies (Kuhn, Hopp \& Els\"{a}sser
1997; Popescu, Hopp \& Els\"{a}sser 1997), consistent with the widely
observed result that galaxies have common voids regardless of Hubble
type (e.g. Thuan, Gott \& Schneider 1987; Babul \& Postman 1990; Mo, McGaugh \&
Bothun 1994). Whether Lyman alpha clouds in voids are
associated with galaxy halos or are a distinct population remains 
controversial (Lanzetta et al. 1995; Morris et al. 1993; see Stocke
2001 for further references). 
Either we are missing a population of galaxies that
reside in the voids or the CDM models produce
too much clustering of matter on small scales (Bode, Ostriker \& Turok 2001). 
The sizes of voids have also been used to place constraints on CDM models.
Blumenthal et al. (1992) and Piran et al. (1993) find that the frequency
and size of voids detected in redshift surveys are difficult to 
reconcile with CDM models unless the Universe is open or that galaxies
do not trace mass on very large scales.

There are several different algorithms in the literature for detecting voids 
(see, for example, Kauffmann
\& Fairall 1991; Kauffmann \& Melott 1992; Ryden 1995; Ryden \& Melott
1996; El-Ad \& Piran 1997 (EP97); Aikio \& M\"{a}h\"{o}nen 1998) but due to
the large diameter of voids (their
characteristic diameter is as large as 50 $h^{-1}$Mpc) 
and the limited volume of most surveys, relatively 
few voids have been objectively detected. 
A search for voids has
so far been made using the 
first slice of the Center for Astrophysics Survey 
(Slezak, de Lapparent \& Bijaoui 1993), the
Southern Sky Redshift Survey (Pellegrini, da Costa \& de Carvalho 1989; 
El-Ad, Piran \& da Costa 1996, EPC96), the IRAS
1.2 Jy Survey (El-Ad, Piran and da Costa 1997, EPC97), 
the Las Campanas Survey (M\"{u}ller et
al. 2000) and the PSCz Survey (Plionis
\& Basilakos 2001). 
Voids were detected in the CfA slice and the
Las Campanas Survey, restricted to two-dimensions due to the small 
declination range of each
data set. Eleven voids with $>$95\% significance have been detected in the
SSRS2 survey (EPC96), twelve voids with $>$95\% significance have been
detected in the IRAS 1.2-Jy Survey (EPC97) and fourteen voids with
volumes larger than $10^3 h^{-3}$ Mpc$^3$ have been detected in the 
PSCz Survey (Plionis \& Basilakos 2001). 
Statistically, voids have also been studied using the 
Void Probability Function 
(VPF, White 1979)
and the Underdensity Probability ($U(R)$, Vogeley et al. 1989), 
which depend on the hierarchy of n-point correlation functions. The VPF is
simply the probability that a randomly selected volume contains no
galaxies. The $U(R)$ measures the frequency of regions with density
contrast $\delta \rho /\rho$ below a threshold. These statistics
reveal statistical information about the void population but do not give details
on specific voids.

A first step in rectifying our relative neglect of voids is to 
quantitatively characterize their properties, as defined by
surveys at different wavelength. Here
we adapt the approach of EP97 in order to search for voids in the PSCz
Survey (Saunders et al. 2000) and the Updated Zwicky Catalog (UZC,
Falco et al. 1999). We describe the surveys in Section \ref{sec:survey}
and describe the algorithm in Section
\ref{sec:algo}. We present our results and draw conclusions in Section
\ref{sec:res} and \ref{sec:conc} respectively.

\section{The Surveys}
\label{sec:survey}

We consider two surveys with different wavelength selection 
to see if the same voids are
detected and if the properties of voids in overlap regions are similar (see
Section \ref{sec:sizevoid}).
We consider different samples from the two surveys to check the
robustness of results. In particular we are able to check the effect
of the wall/field criteria on the detection of voids 
(see Section \ref{sec:resfield}) and by how much the
volume of voids is restricted by the sample depth
(see Section \ref{sec:resedge}).

\subsection{The Updated Zwicky Catalog}
\label{sec:UZC}

The Updated Zwicky Catalog (Falco et al. 1999) includes a re-analysis of data
taken from the Zwicky Catalog and Center for Astrophysics redshift survey to
$M_{Zwicky} \lsim$ 15.5
(Zwicky et. al 1961-1968; Geller \& Huchra 1989; Huchra et al. 1990;
Huchra, Geller \& Corwin 1995; Huchra, Vogeley \& Geller 1998
) together with new spectroscopic redshifts for some galaxies
and coordinates from the 
digitized POSS-II plates. Improvements over the previous catalogs 
include estimates of the accuracy of the CfA redshifts and 
 uniformly accurate coordinates at the $< 2^{\prime \prime}$
level.
The UZC contains a total of
19,369 galaxies. Of the objects with m$_{Zwicky} \lsim 15.5$, 96\% 
have measured redshifts, giving a total number of 18,633 objects. The
catalog covers two main survey regions; $20^h < \alpha_{1950} < 4^h$
and $8^h < \alpha_{1950} < 17^h$ both with $-2.5^{\circ} <
\delta_{1950} < 50^{\circ}$.

We consider three different samples from the catalog. We
construct a volume-limited sample with $z_{\rm max}$=0.025 
corresponding to a depth of 74$h^{-1}$Mpc, for an Einstein-de Sitter
cosmology. The absolute magnitude of each galaxy is estimated to be
\begin{equation}
M = m_{Zwicky} - 25 - 5{\rm log}[r(1 + z)] - 3z
\end{equation}
and this sample is limited to $M_{lim}= -18.96$.
Previous analyses of the CfA or UZC samples have removed 
areas in the South Galactic Cap region with extinction larger than 
$\Delta m \gsim 0.3$. We do not apply this cut
as it severely restricts the volume of the survey. Instead, we compare the voids
we find in the UZC with those we find in PSCz to examine if the paucity of
galaxies in these regions of the catalog causes us to falsely detect voids. 

This volume-limited sample 
contains 3518 galaxies, which is the largest number of
galaxies in a volume limited sample that can be constructed from the
catalog. Application of the wall/field criteria 
(defined in Section \ref{sec:wallfield}) to this sample 
yields 3240 wall galaxies and 278 field galaxies. We also 
consider the same volume limited sample without applying
the wall/field criteria to see how the voids
are affected when we insist they are completely empty. We also consider a
sample that extends an extra 20$h^{-1}$Mpc in depth to estimate the 
impact of the
survey boundary on the estimated sizes of voids.

\subsection{The PSCz Survey}
\label{sec:PSCz}

Objects in the PSCz Survey were selected from the IRAS Point Source
Catalog (Beichman et al. 1984) which is a catalog of detections down to
a flux limit of 0.6 Jy taken with
the IRAS (Infra-Red Astronomical Satellite). Targets for the PSCz
survey were chosen to maximise the sky coverage but at the same time
other considerations such as completeness, flux uniformity, having a
well defined sample area and redshift range were taken into
account. (See Saunders et al. 2000 and references therein for a more
complete description of target selection criteria.)

The PSCz survey covers 84\% of the sky. Areas with incomplete
IRAS data and high optical extinction, such as the plane of the
Galaxy, are not included in the Survey.
The PSCz survey contains 15,411 galaxies. Of order 8,000
objects had known redshifts in the literature (Saunders et
al. 2000). Follow-up observations were carried out to obtain
the remaining redshifts. The median redshift of the survey is 0.028
corresponding to a comoving depth of $\sim 80 h^{-1}$Mpc assuming
an EdS cosmology.

In a flux limited
sample the number density of detectable objects decreases with
distance. Thus, we are more likely to detect voids towards the
edge of the survey because the average galaxy-galaxy separation
increases. Therefore we restrict the depth of the sample to the
median redshift, 0.028, which closely matches the depth of the UZC sample. 
As IR selected surveys are sparse compared to optically selected surveys,
we keep all the galaxies in the sample, {\it i.e.} we do not construct a 
true volume-limited catalog. We find that void properties do not vary 
with depth over the range we consider, thus the void finding algorithm
is robust with respect to small changes in the selection function.
The average density of objects in this 
PSCz sample is comparable to the density of objects in the 
volume-limited UZC sample if
we do not apply the volume limited criteria to PSCz.
This allows us to compare voids found in the two surveys at 
similar sampling density.

As for the UZC, we consider this sample in two ways, 
one where we we apply the wall/field
galaxy criteria (defined in Section \ref{sec:wallfield}) 
and one where we do not (and so all galaxies are wall
galaxies). This sample contains 8425 galaxies (7743 wall galaxies and 
682 field galaxies)
A third sample is constructed in
which we extend the sample by 20$h^{-1}$Mpc. Again, this is to examine by how
much we underestimate the volume of voids when we impose our cut in
depth.

\section{The Void Finding Algorithm}
\label{sec:algo}

\subsection{Outline}

Here we outline the steps involved in our void
finding algorithm. Each of the steps are discussed in detail in the
subsequent sections.  The void finding algorithm we adopt is very
similar to the method used by El-Ad \& Piran ({\tt Void Finder} EP97,
EPC97). 

Tests with toy model distributions and results of N-body simulations
indicate that this algorithm is robust in locating voids and measuring
their sizes.  The method allows for ``field galaxies'' in the voids,
which are removed before void detection. This step ensures that void
detection is not dominated by shot noise of the few galaxies in these
large underdensities. Setting a minimum void size prevent voids from
percolating through small gaps in the dense wall-like or filamentary
structures. This minimum void size is also set so that the voids are
statistically significant in comparison with randomly occurring holes
in a Poisson distribution with the same average density of objects. As
long as the sampling density is high enough that the overdense
structure are well-defined (no gaps comparable to or larger than the
minimum void threshold), then void detection is not highly sensitive
to the sampling density.

The steps of the void finding algorithm are as follows:
\begin{itemize} 
\item Classification of galaxies as wall or field galaxies, Section \ref{sec:wallfield}
\item Detecting empty cells in the distribution of wall galaxies, Section \ref{sec:cells}
\item Growth of the maximal empty spheres, Section \ref{sec:spheres}
\item Classification of the unique voids, Section \ref{sec:classvoids} 
\item Enhancement of the void volume, Section \ref{sec:volume}
\end{itemize}

\subsection{Wall and Field Galaxies}
\label{sec:wallfield}

The first stage in identifying voids is to determine which galaxies
are classed as field galaxies and which are wall galaxies. Here we
follow the method of EP97. First we calculate the mean distance, {\it
d}, to the $n^{\rm th}$ galaxy and the standard deviation on this
value, $\sigma$. We then specify a length {\it l$_n$} such that any
galaxy that does not have $n$ neighbours within a sphere of radius
{\it l$_n$} is classified as a field galaxy. The value of {\it l$_n$}
we adopt is given by {\it l$_n$} = {\it d} + 1.5$\sigma$ and we set
$n$=3. For our main samples, the values of {\it d} and {\it l$_3$}
that we obtain for the PSCz survey are 3.42$h^{-1}$Mpc and
7.19$h^{-1}$Mpc respectively while for the UZC we obtain 3.46$h^{-1}$
and 6.52$h^{-1}$Mpc. This choice of sphere size means that field galaxies
are found in regions with density contrast, $\delta \rho/\rho$, of less
than -0.5. For {\it n}=3 $\sim$10\% of
galaxies are classified as field galaxies in both surveys. If $n$=0 then all
the galaxies are field galaxies and the survey is just one large void
(or two because of the disjoint geometry of both the PSCz and UZC). If
$n$ is higher than 3, then the proportion of galaxies classified as
field galaxies drops rapidly. We compare results for our main sample to 
an analysis of a
sample where all the galaxies are classed as wall galaxies to see what
effect the exclusion of field galaxies has on the voids we detect.

We also have to consider what effect the edges of the survey have on our
wall/field criteria. Galaxies close to the edges may appear isolated
and be classified as field galaxies but they may be part of large
groups just outside the geometry of the sample 
under consideration. 
This could lead to a slight overestimation of the
size of the voids near the survey boundary 
because a maximal sphere in the vicinity of an erroneously classified
edge galaxy can be grown to the survey boundary rather than being 
bounded by that galaxy.
As these wrongly classified
field galaxies are only those close to the survey edge, 
the maximal spheres cannot
be grown much larger and hence the void sizes should not be overestimated
by a large amount. There is no indication that voids near the boundaries are larger than those interior to the surveys.

Similarly, we may overestimate the values of $\delta \rho/\rho$ for 
voids that lie close to the edges as the wrongly classified field 
galaxies may lie within a void. However, on inspection, few of these
galaxies actually lie in voids due to the spherical nature of voids
and the proximity of the survey boundary. 

\begin{figure*} 
\begin{centering}
\begin{tabular}{ccc}
{\epsfxsize=5truecm \epsfysize=5truecm \epsfbox[35 170 550 675]{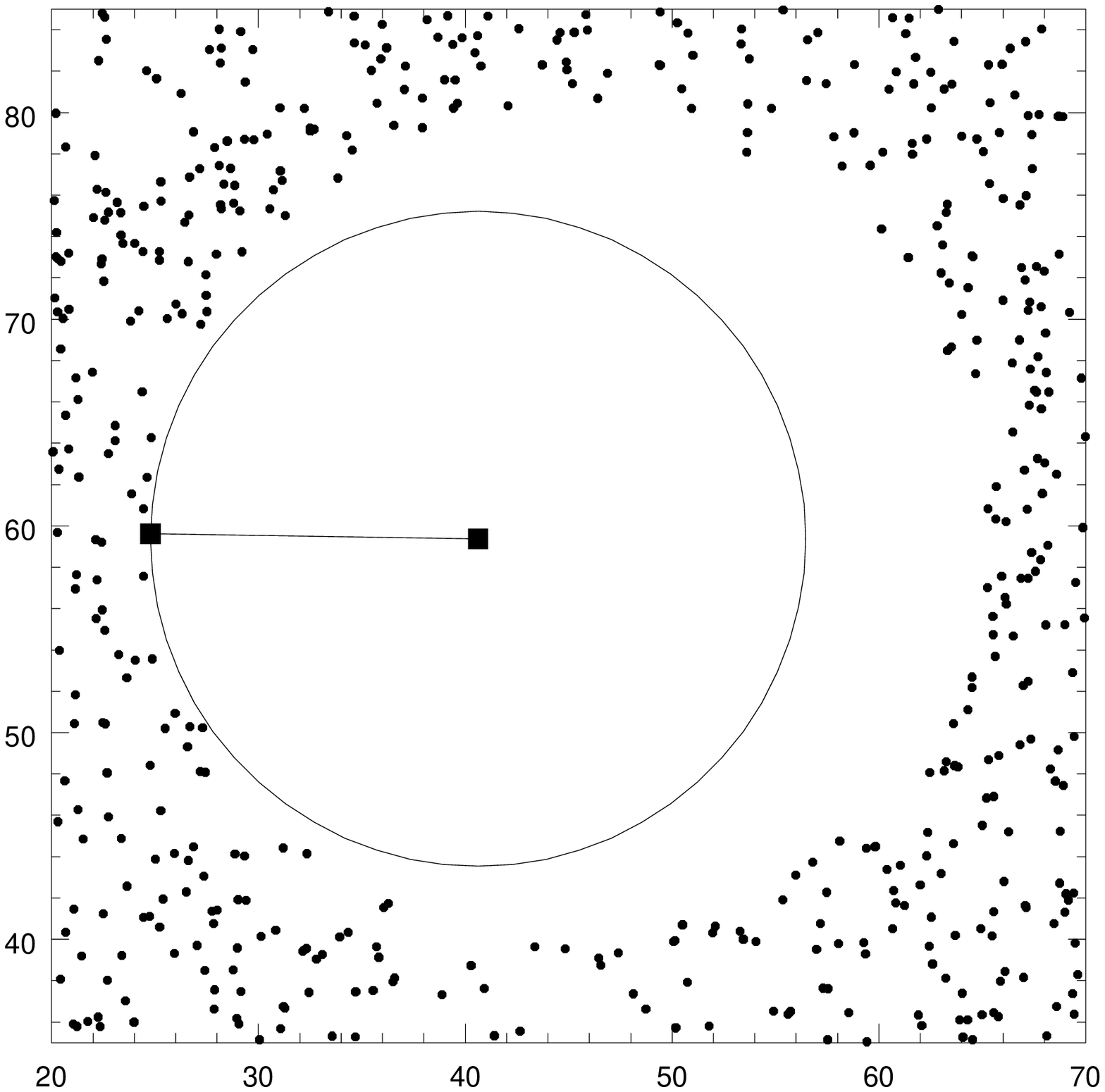}} &
{\epsfxsize=5truecm \epsfysize=5truecm \epsfbox[35 170 550 675]{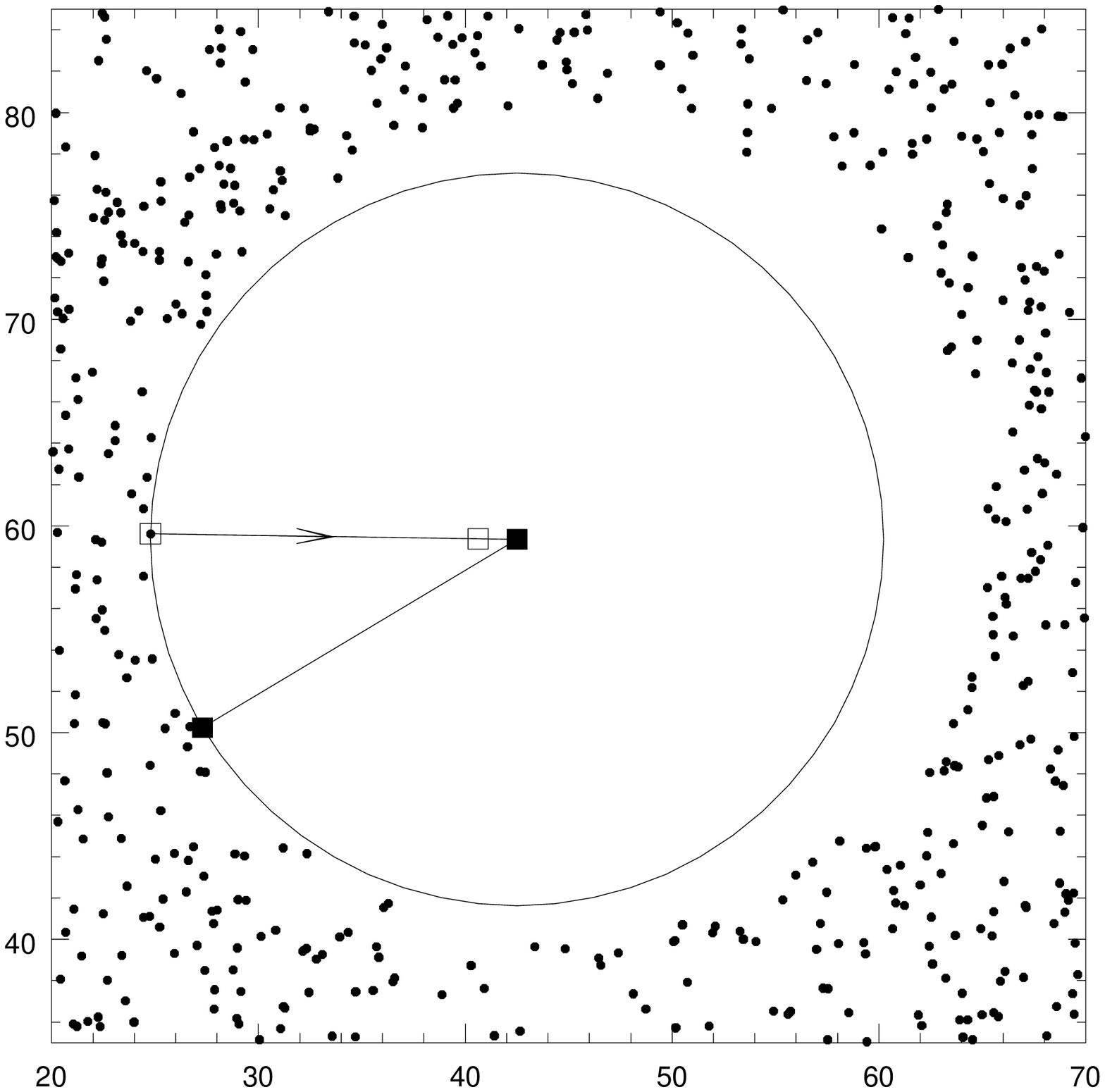}} &
{\epsfxsize=5truecm \epsfysize=5truecm \epsfbox[35 170 550 675]{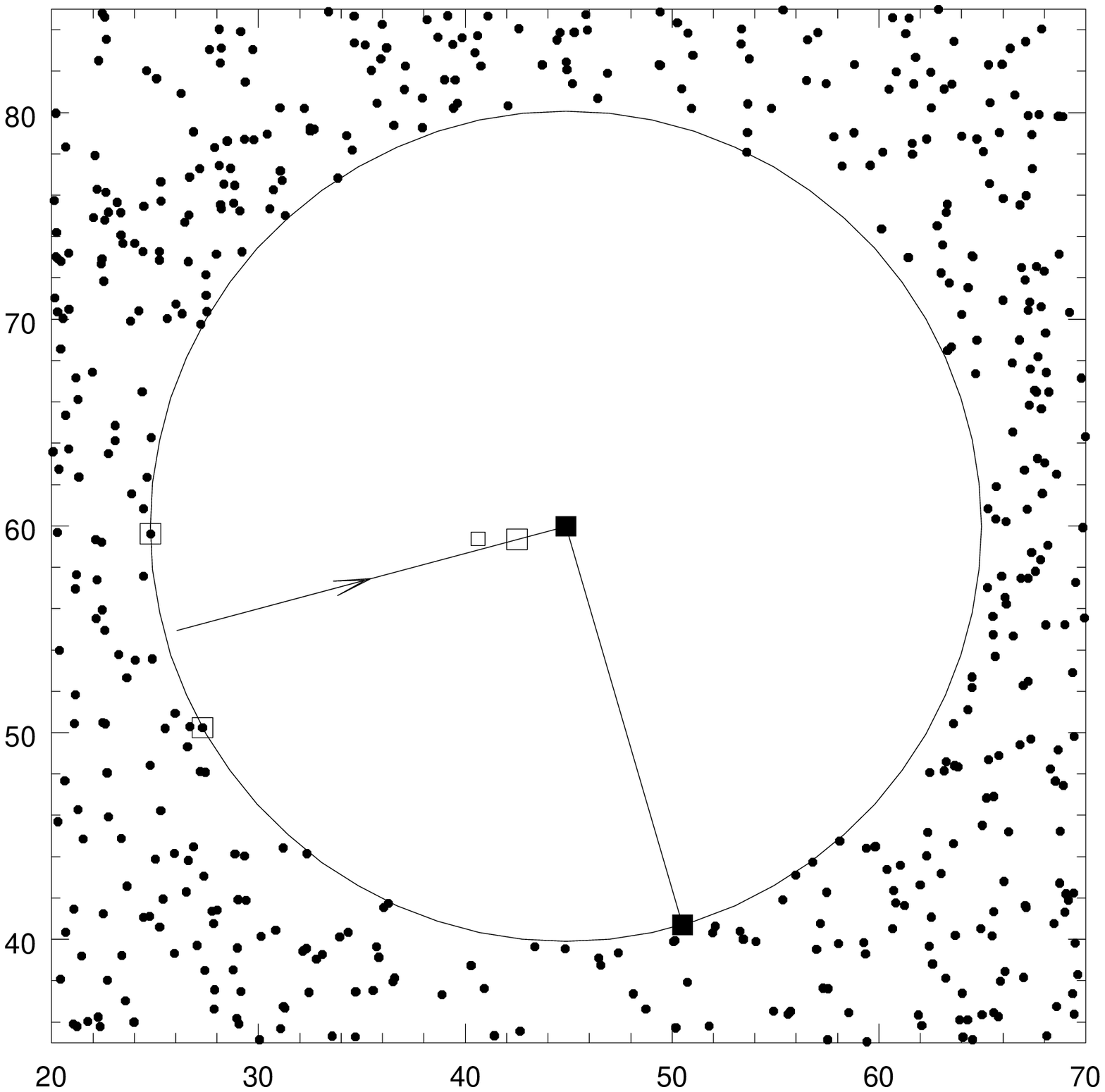}} \\
\end{tabular}
\caption{ The void finder technique. The method is discussed in more
detail in the text. Step 1, we identify an empty cell within a
void. From that cell we find the nearest wall galaxy and grow the sphere to
that radius. Step 2, we move away from the first galaxy, growing the
sphere until the second galaxy is on the surface. Step 3, we connect the
sphere center to the point which bisects the two detected galaxies and
move the center along this vector until a third galaxy is on the
radius. In 2D this is where we stop. In 3D, the final stage is to move
the sphere center out of the plane defined by the first three
galaxies. The open squares in the diagrams are previous void centers
and earlier detected galaxies. The solid squares show the
current void center and the galaxy detected at each stage. 
The points in this two dimensional example 
are randomly generated in a box with a circular area carved out as a
demonstration of the technique.}
\label{fig:findvoids}
\end{centering}
\end{figure*}

\subsection{Detecting the Empty Cells}
\label{sec:cells}

The next step is to place the galaxies (either all the
galaxies or just the wall galaxies when we apply the wall/field
criteria) onto a three dimensional grid and count the number of
galaxies in each cell. Cells that lie at a greater distance than the
maximum radius of the sample under consideration are not considered
further. The fineness of the grid defines the minimum size void that
can be detected. If each grid cell has length l$_{\rm cell}$, then
voids with radius r = $\sqrt{3}$l$_{\rm cell}$ will be found. We
fix l$_{\rm cell}$ to be the same as {\it l$_3$}.

\subsection{Growth of the Maximal Spheres}
\label{sec:spheres}

Each empty grid cell (we will refer to these as holes) is considered to be part
of a
possible void. Our method finds the maximal sphere that can be drawn in
the void, starting from the empty cell, and is shown
schematically in Figure \ref{fig:findvoids}. From the center of
the empty cell, we grow a sphere, increasing its radius
until we find a galaxy that is
just on the edge of it. We then find the vector that connects this
galaxy with the center of the hole and move the center of the
spherical hole along this direction, away from the first galaxy,
growing the radius to keep the first galaxy on its surface, until a
second galaxy is also on the surface. We next find the vector that
bisects the line joining the two galaxies and move the hole in this
direction until a third galaxy is found, as before. If voids are being
detected in two dimensional data, the process stops here. However, we
are considering three dimensional data so the final step is to grow
the hole out of the plane formed by the first three galaxies. We
insist that our voids lie completely in the survey {\it i.e.} they are
contained within the volume specified by our survey boundaries and
sample depth. At this stage we keep track of all the holes with radii
larger than the value of the search radius used to classify field and
wall galaxies, $l_{3}$.

Our method is somewhat redundant. We specify the minimum
size of our holes that form voids but many voids are far larger than
this and more than one empty cell may lie within the void
region, thus we grow some voids more than once. 
However, it does guarantee that we find all the holes that can
contribute to the voids volume so the redundancy is eliminated when
we calculate the volume of the voids (see Section \ref{sec:volume}).

\subsection{Classification of Unique Voids}
\label{sec:classvoids}

Finding the holes is a robust process. Deciding which of those
holes are unique voids requires more care. 
Our definition of a void is slightly different
from that of EP97. First we sort the holes by radius, the largest
first. The largest hole found is automatically a void.
We define a fractional overlap parameter, $\eta_1$.
If the second void overlaps the first by more than $\eta_1$ in volume,
then we say it is a member of the first void rather than a new
void. If not then it forms a separate void. We then check the third
hole to see if it overlaps either of the previous voids by $\eta_1$ in
volume. If it does, we add it to that void; if not, we form a new
void. If it overlaps two voids by more than $\eta_1$ then we
reject the hole as it links two larger, independent voids.
We continue like this for all holes with radii larger than
10$h^{-1}$Mpc.

We have investigated how much holes should overlap and yet be
considered separate voids. If the overlap threshold is high, say $\eta_1$=90\%,
then most of the holes are considered unique voids and the same empty 
region may be contained in more than one void. If the overlap fraction is too
low, then voids that barely overlap are considered two parts of the same 
void. This means that if there are any small gaps in the walls,
two spheres could be joined together to form a single void which, under
visual inspection, obviously consists of two voids.

\begin{figure} 
\begin{centering}
\begin{tabular}{c}
{\epsfxsize=5truecm \epsfysize=5truecm \epsfbox[35 170 550 675]{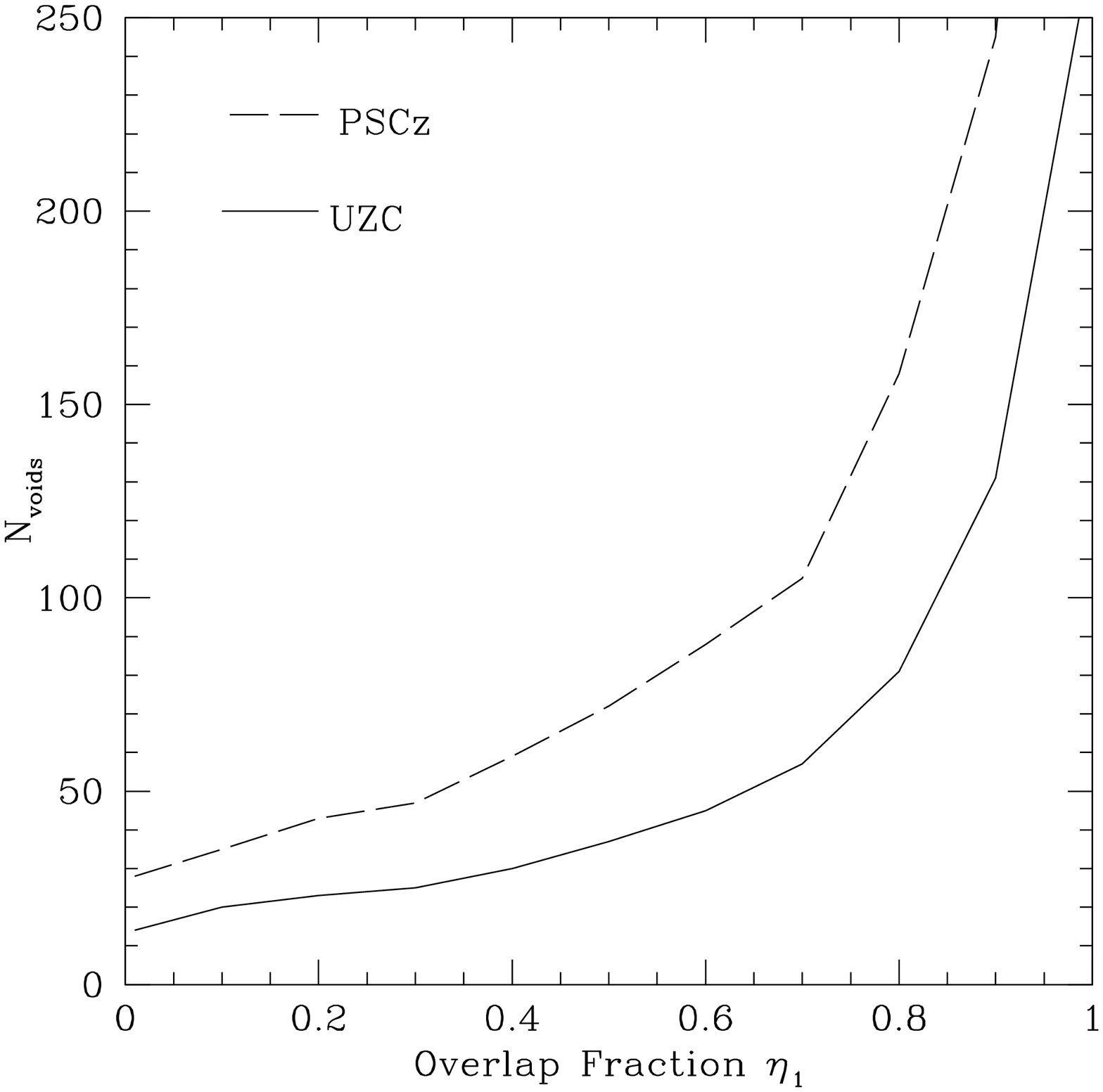}} \\
\end{tabular}
\caption{The number of voids in each survey as a function of
overlap fraction, $\eta_1$, allowed between distinct voids 
(dashed line PSCz, solid line UZC). If
spheres are allowed to almost completely overlap and still be
classified as individual voids, then we detect too many voids in each
survey. We adopt a value of $\eta_1 =$10\%.}
\label{fig:groups}
\end{centering}
\end{figure}

In figure \ref{fig:groups} we show the
number of voids we find for the PSCz (dashed line) and the UZC (solid
line) as we vary the percentage overlap. 
The values seem to flatten off for $\eta_1 <$30\%. We fix the value 
of $\eta_1$ to be 10\%. 
Visually, using our criteria of $\eta_1$=10\%, voids appear to be
distinct regions of the survey. 
For a void with a radius of 10$h^{-1}$Mpc and with an overlap
region of $\eta_1$=10\%, gaps in the walls would have to be larger
than 9$h^{-1}$Mpc, which is larger than the wall/field galaxy criteria
and larger than the average galaxy-galaxy separation in both surveys.
Gaps of this size in the walls are rare so distinct voids are not 
connected. Therefore, if a hole with radii larger than 10$h^{-1}$Mpc
overlaps a larger void by more than 10\% of its volume, we merge the
hole into that void. If $\eta_1$ is less than 10\% of the
smaller void's volume, we deem the sphere to be a distinct
void and if the hole overlaps two larger voids by 10\% we reject it.

We set a threshold of 10$h^{-1}$Mpc for the minimum size of voids. 
This threshold is larger than the
search radius for defining field galaxies, $l_3$. This helps ensure
that we do not identify gaps in the walls as voids. It is also the
value at which the significance of detecting voids in the both the
PSCz and UZC catalog drops below 95\%, discussed in Section 
\ref{sec:conf}. 

\subsection{Enhancement of the Void's Volume}
\label{sec:volume}

We next enhance the volume of each void. We consider the holes that
have radii less than the threshold of 10$h^{-1}$Mpc but greater than
the radius for the wall/field galaxy criterion, {\it l$_3$}, 
(in the case where we do
not make the wall/field galaxy cut we still use {\it l$_3$} as the minimum 
sphere radius). Any hole that
overlaps the maximal void sphere (as grown in Section \ref{sec:spheres})
by $\eta_2 > $50\% of the smaller hole's volume
is also considered part of the larger void. If the hole overlaps with
more than one void then it is not added to either of the voids as this
would link two voids together that we wish to keep separate. If the
hole is isolated it cannot be classed as a separate void as it is
smaller than the threshold we use for void classification. The choice
of $\eta_2=$50\% is somewhat arbitrary but it fills the
void volume without changing the overall spherical shape of the voids.
We have compared the volumes of voids found using different values of 
$\eta_2$. For values of $\eta_2$ in the range 20-70\%, the volumes
of voids are robust to within 20\%.

We compute the volume of each void by Monte Carlo integration, i.e. we
embed it in a box that is larger than the void and generate many
random particles within the box and count how many lie within one of
the holes that make up the void. The ratio of the number of randoms in
the void to the total number of randoms placed in the box multiplied
by the volume of the box gives an estimate of the volume of the
void. We could, in principle, calculate the volume of the void
analytically as it is just formed by a series of overlapping
spheres. However, if a void consists of more that three overlapping
holes, the solution is rather complicated.

\begin{figure} 
\begin{centering}
\begin{tabular}{c}
{\epsfxsize=5truecm \epsfysize=5truecm \epsfbox[20 170 570 690]{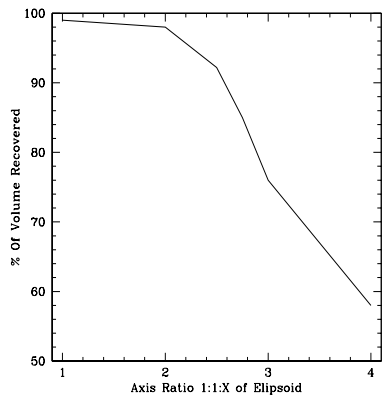}} \\
\end{tabular}
\caption{The accuracy to which we recover the volume of an
ellipsoid with axis ratios 1:1:X. If the void is spherical then we
recover 100\% of the volume of the sphere. The more elongated the
void, the smaller the fraction of the volume we recover
when requiring that the maximum sphere size be larger than the
wall/field criteria, $l_3$. We only show the results for ellipses with
two axes equal and the third different as this is the worst case
scenario. If the axis are in the ratio 1:1.5:2 we recover 90\% of the
volume and if they are in the ratio 1:2:3 we recover 70\% of the
volume.}
\label{fig:axis}
\end{centering}
\end{figure}

We are likely to underestimate the volumes of voids as we only
consider holes that have radius greater than the search radius,
$l_3$. If voids are highly elliptical then we will not detect the
volume at the `corners' of the ellipse. We test this by generating
data containing mock voids of known elliptical shape. We take the
average volume of a void, taken to be 15,000$h^{-3}$Mpc$^3$, and
generate ellipsoids with this volume that have axis ratios 1:1:X
(i.e. if X is 1 then the void is spherical with radius
15.3$h^{-1}$Mpc). We then run the simulated data through our void
finding algorithm and compare the volume obtained with the known
volume of the void. The results are shown in Figure \ref{fig:axis}. If
voids are spherical in shape then we recover 100\% of its volume. The
more elongated it becomes, the less of the volume we detect. This
becomes significant if one of the axes is elongated by more than a
factor of 2.5 than the others. If the axis are in the ratio 1:1.5:2
then we can recover 90\% of the volume and if they are in the ratio
1:2:3 we can recover 70\% of the volume. As voids are seen to be
roughly spherical in nature we should be able to detect most of a
void's volume.

Finally in the case where we specify wall and field galaxies, we count
the number of field galaxies that lie in each void to determine the
underdensity of the void.

\begin{figure*} 
\begin{centering}
\begin{tabular}{c}
{\epsfxsize=16truecm \epsfysize=16truecm \epsfbox[10 150 570 690]{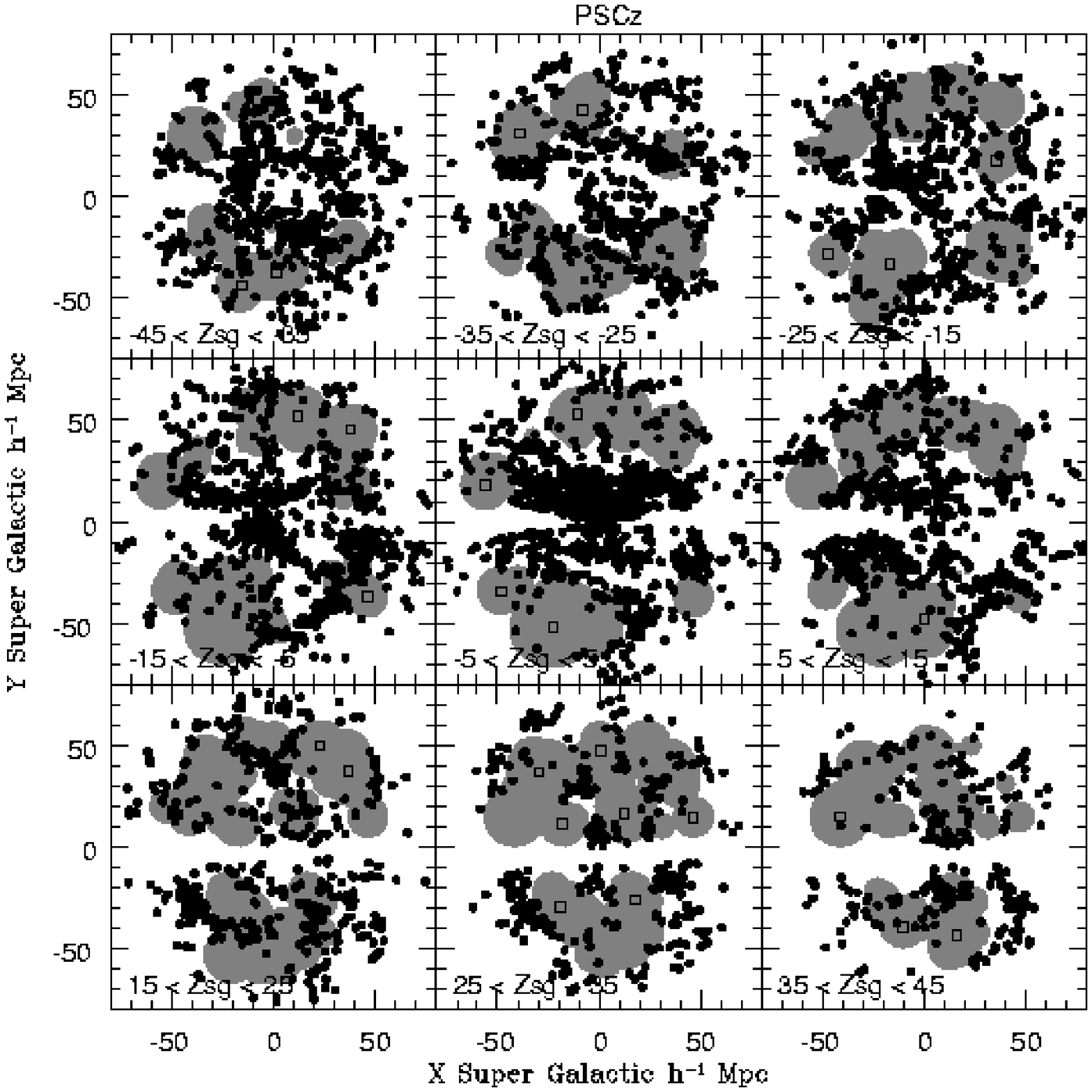}}
\end{tabular}
\caption{Voids in
the PSCz Survey. We show the supergalactic coordinates (X,Y) for
different values of Z. Each panel shows a 10$h^{-1}$Mpc 
slice starting at -45$h^{-1}$Mpc$<$Z$<$-35$h^{-1}$Mpc top left to 
35$h^{-1}$Mpc$<$Z$<$45$h^{-1}$Mpc bottom right. The
shaded regions are the voids. The filled points are the wall
galaxies and the open squares show the void centers (as the voids have 
radii $> 10 h^{-1}$ Mpc they spread over more than one panel and thus some 
shaded areas do not contain a void center). No wall galaxies
are found in the voids. Field galaxies are not shown.}
\label{fig:psczvoids}
\end{centering}
\end{figure*}

\begin{figure*} 
\begin{centering}
\begin{tabular}{c}
{\epsfxsize=16truecm \epsfysize=16truecm \epsfbox[10 150 570 690]{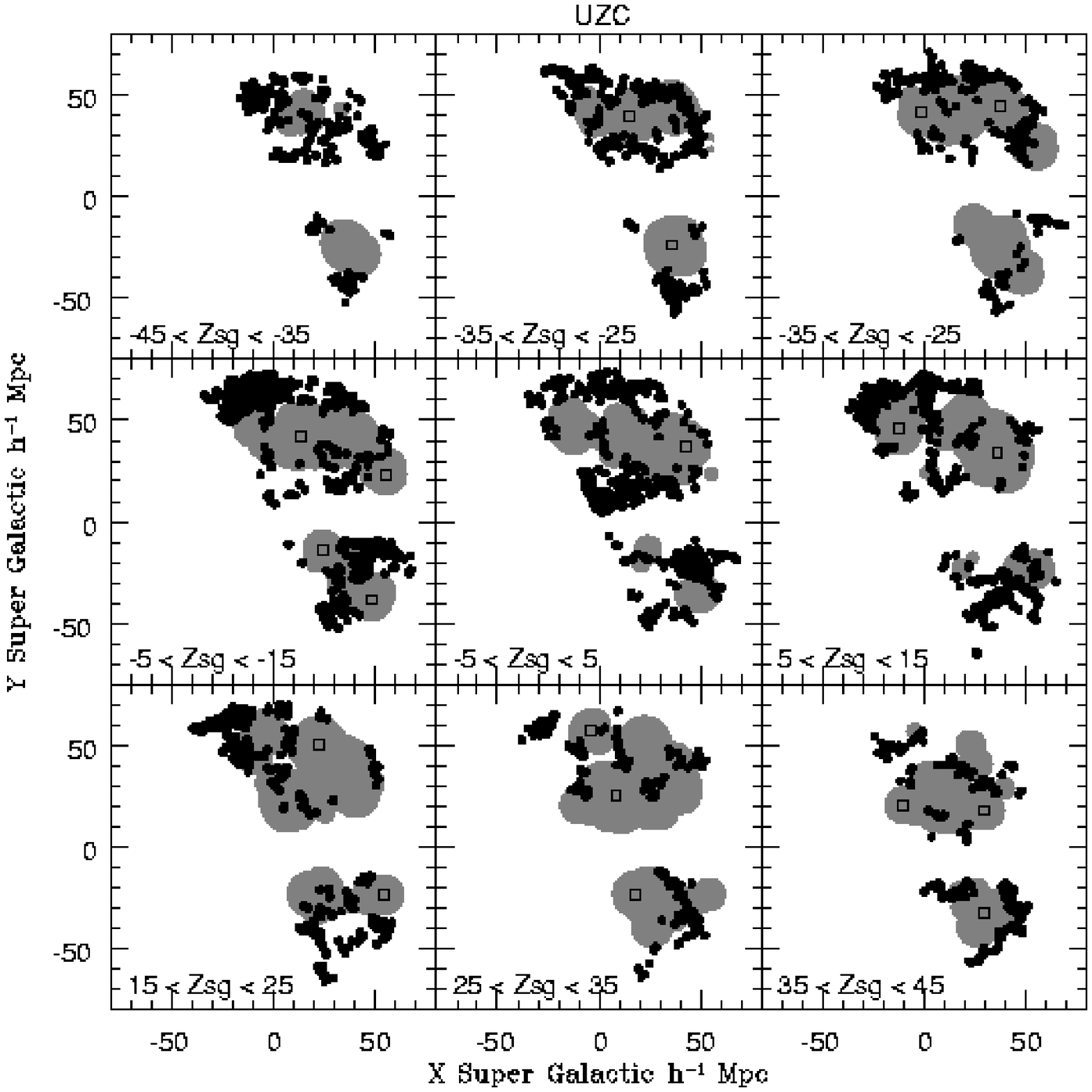}}
\end{tabular}
\caption{ Voids in
the UZC Survey. We show the supergalactic coordinates (X,Y) for
different values of Z. Each panel shows a 10$h^{-1}$Mpc 
slice starting at -45$h^{-1}$Mpc$<$Z$<$-35$h^{-1}$Mpc top left to 
35$h^{-1}$Mpc$<$Z$<$45$h^{-1}$Mpc bottom right. The
shaded regions are the voids. The filled points are the wall
galaxies and the open squares show the void centers (as the voids have 
radii $> 10 h^{-1}$ Mpc they spread over more than one panel and thus some 
shaded areas do not contain a void center). No wall galaxies
are found in the voids. Field galaxies are not shown.}
\label{fig:cfavoids}
\end{centering}
\end{figure*}

\begin{figure*} 
\begin{centering}
\begin{tabular}{ccc}
{\epsfxsize=5truecm \epsfysize=5truecm \epsfbox[35 170 550 675]{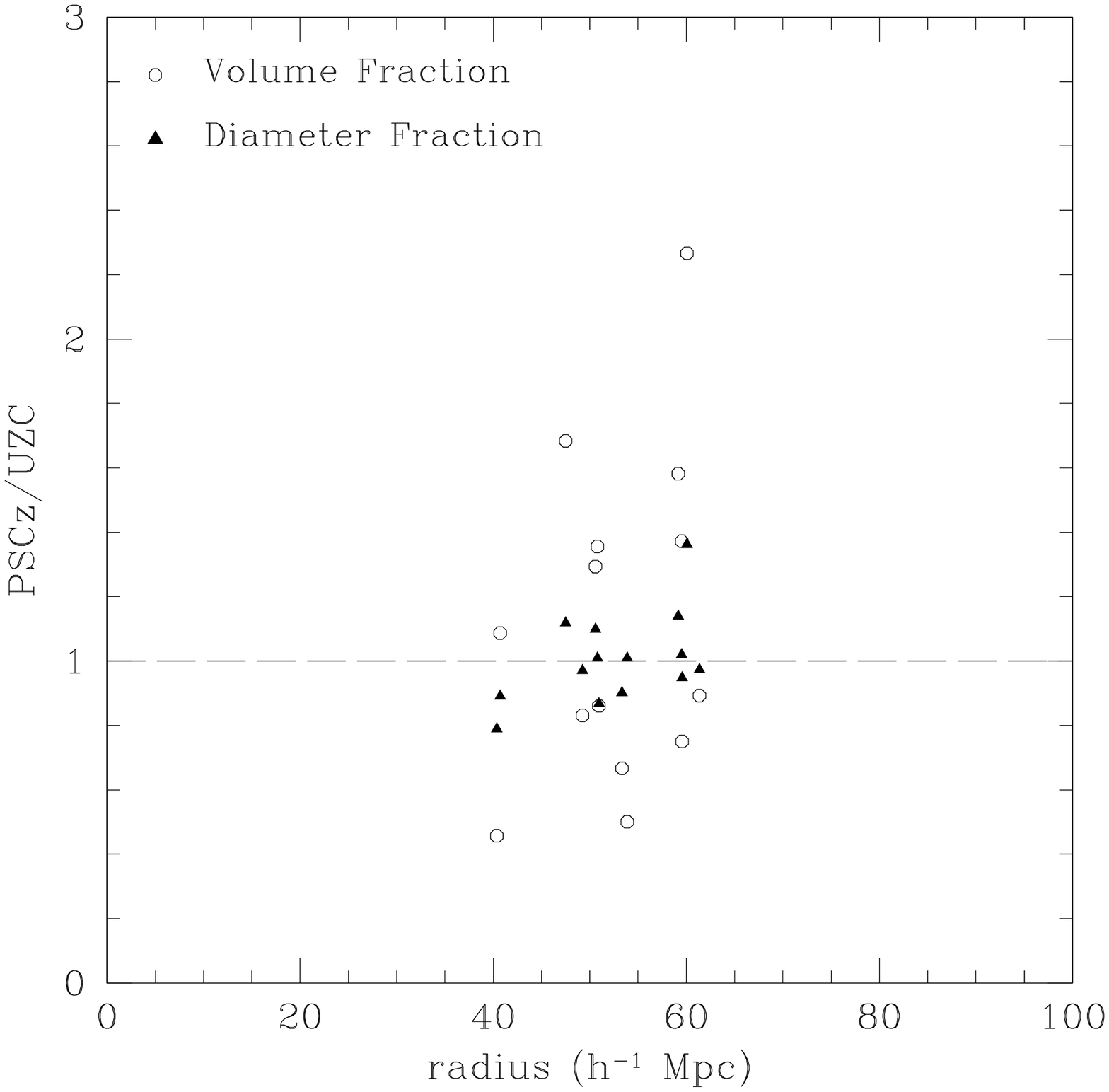}} &
{\epsfxsize=5truecm \epsfysize=5truecm \epsfbox[35 170 550 675]{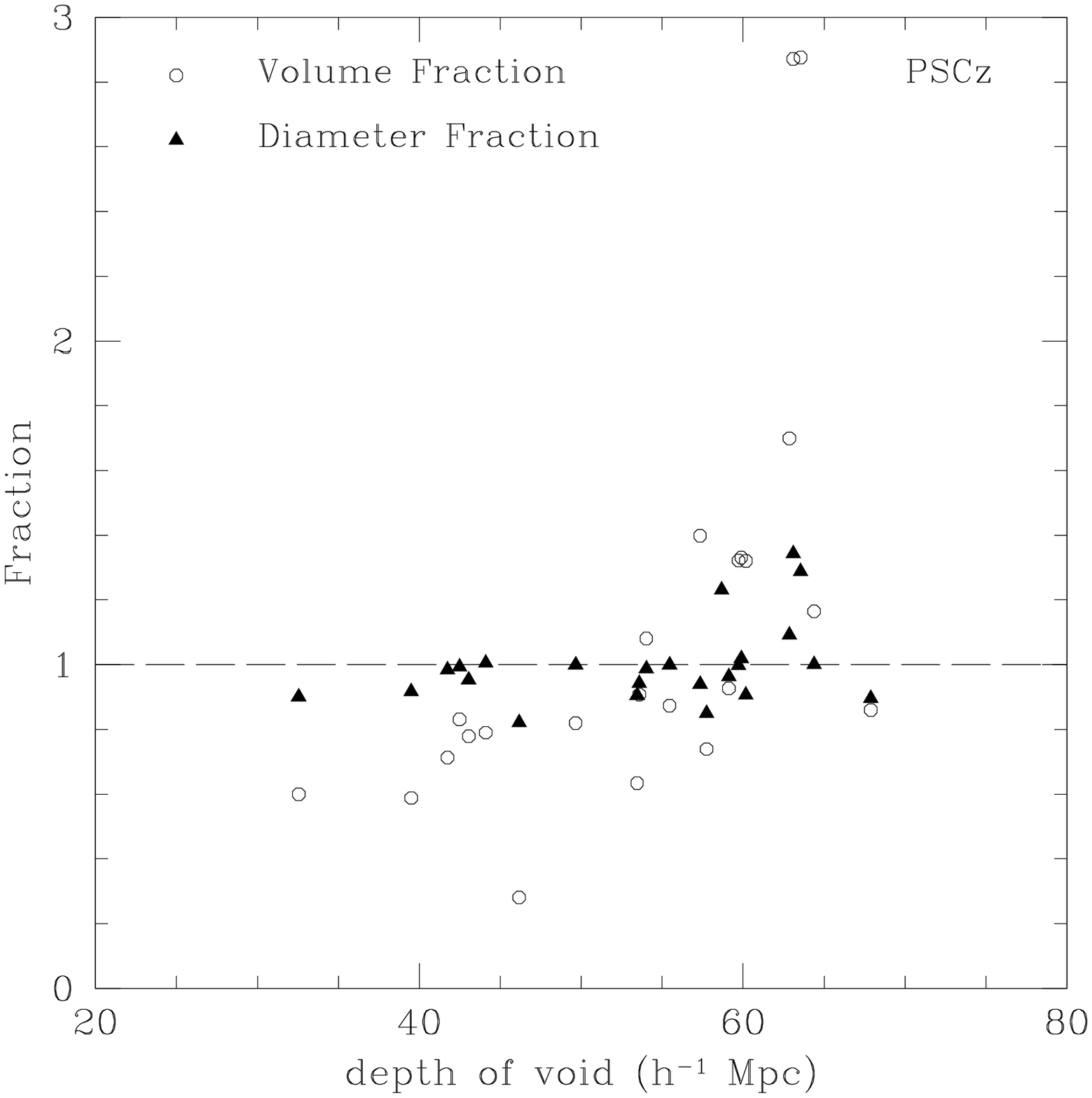}} &
{\epsfxsize=5truecm \epsfysize=5truecm \epsfbox[35 170 550 675]{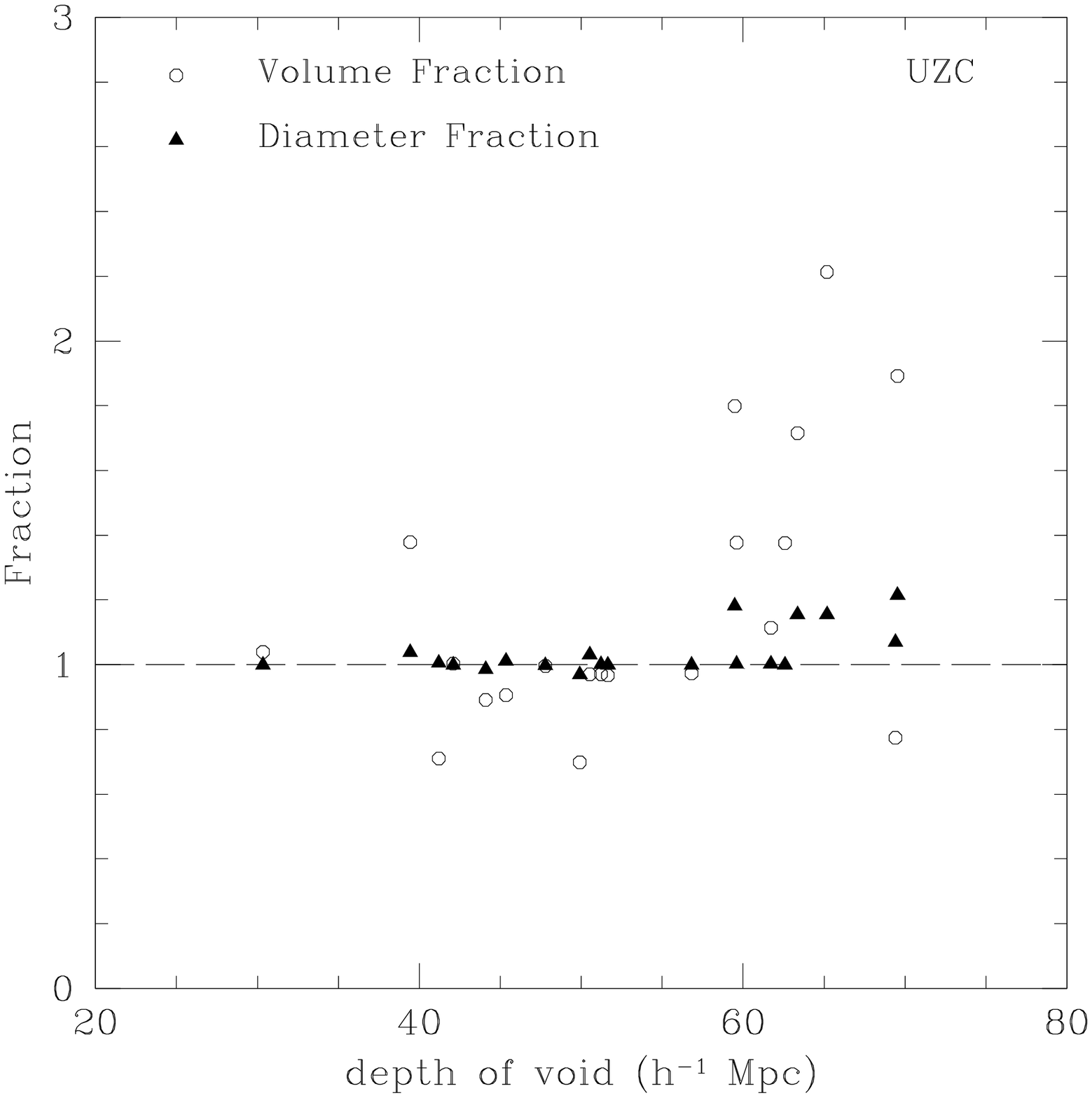}} \\
\end{tabular}
\caption{The left hand plot shows a comparison between the radii of the maximal sphere and the volume of voids detected in both the PSCz and UZC catalogue. The radii of the voids detected in the two surveys agree to with in 20\%, indicating that sample selection plays little effect on the diameters of voids. The center and right hand plot give an indication of the amount we underestimate the radii of the maximal spheres (triangles) and the volumes (circles) of voids by insisting they lie entirely in the volume of the survey as a function of depth for the PSCz survey (center plot) and the UZC (right hand plot). Voids that lie closer to the depth boundary naturally suffer more from underestimation problems but even so, we estimate the radii are accurate to within 30\%. }
\label{fig:underest}
\end{centering}
\end{figure*}

\section{Results}
\label{sec:res}

Voids in the main PSCz and UZC
sample are shown in Figure \ref{fig:psczvoids} (PSCz) and
\ref{fig:cfavoids} (UZC). In both cases, the points represent the wall
galaxies and the shadings represent areas that are covered by
voids. Squares indicate the three dimensional center of each void.
Projecting 3-D data onto a 2-D page makes the voids appear to
contain galaxies but in reality voids are free from wall galaxies (but
may contain field galaxies which are not shown). We find 35 voids 
with r$ > 10 h^{-1}$Mpc in the 
PSCz survey and 19 voids with r$ > 10 h^{-1}$Mpc in the UZC.

\subsection{Significance of the Voids}
\label{sec:conf}

We use the method of EPC97 to assess the significance of
voids. The confidence level with which we detect a void is given by
\begin{equation}
p(r) = 1 - \frac{N_{\rm Poisson}(r)}{N_{\rm Survey}(r)}
\end{equation}
where $N_{\rm Survey}(r)$ is the number of voids in the sample
under consideration and $N_{\rm Poisson}$ is the number of
voids found in Poisson realisations of the sample. These have the same 
number density of points and the same radial and angular selection functions
of the sample but the points are unclustered. The closer $p(r)$ is to 1,
the less likely a void could occur in a random distribution. 

With a threshold of 10$h^{-1}$ Mpc for the void radius, we find an 
average of 1.9  voids in the mock PSCz realisations and an average
of 1.1 voids in the UZC realisations. Thus the voids we detect
in both the PSCz and UZC with radius $>$10$h^{-1}$ Mpc 
are significant at the 95\% confidence level, which is why
we set this minimum size threshold.

\subsection{Sizes of the Voids}
\label{sec:sizevoid}

The largest hole in the PSCz survey has a diameter of 35.7$h^{-1}$Mpc,
whereas in the UZC survey the largest hole has a diameter of
29.3$h^{-1}$Mpc. The wider angle covered by the PSCz survey
allows somewhat larger voids to be detected. The average void in the PSCz
survey has a maximal sphere diameter of 24.8$\pm 3.3 h^{-1}$Mpc and an
effective diameter of 29.8$\pm 3.5 h^{-1}$Mpc (we find the effective diameter
by calculating the diameter of a sphere that would match the volume of the void)
whereas the average void
in the UZC survey has a slightly smaller maximal sphere diameter of
23.6$\pm 2.7 h^{-1}$Mpc and an effective diameter of 29.2$\pm 2.9 h^{-1}$Mpc. 
The average void size obviously depends on the minimum size threshold
of r$>$10$h^{-1}$ Mpc. Raising or lowering this threshold raises or lowers the
average void size. 
The maximal spheres typically fill 60\% of the void's volume, found by
dividing the volume of the void by the volume of the maximal sphere. There 
does not appear to be any trend with void size in this filling fraction.  

We compare the voids detected in both
the PSCz and the UZC. We are able to
match 15 out of the 19 voids in the UZC to voids in the PSCz. 
Two (10 and 16 in the UZC list) 
of the voids detected in the UZC form one void in the PSCz
survey. 
Void 19 appears to be a case where the maximal
sphere in the PSCz survey lies just below the threshold of 10$h^{-1}$Mpc
and so is not classed as a void in the PSCz. 
Voids 12 and 18 (marked with an asterisk in table \ref{tab:uzcvoids}) 
are found in areas with high extinction in the South Galactic Cap.
There are few galaxies detected in the UZC in these regions so 
voids can be grown. As extinction is less of a problem in the IR, 
relatively more galaxies are detected in the PSCz sample and 
voids are not detected.
Thus, these particular voids are suspect and we exclude them from 
calculation of the statistical properties of voids.

For the 15 voids that
we do detect in both of the surveys, the diameter of the maximal spheres
and the volumes we detect are very similar. In figure
\ref{fig:underest} we divide the diameters and volumes of the voids
detected in the PSCz by the same values found from the UZC and find
values fairly close to 1.

\subsection{Density of Voids}
\label{sec:voidrho}

In the case where we differentiate between wall and field galaxies, we
can determine the density contrast of voids. We
find that voids have a typical density contrast of $\delta \rho / \rho$ =
-0.92$\pm0.03$ in the case of the PSCz survey and -0.96$\pm0.01$ 
in the case of the UZC 
survey. These values are very low; even with 10\% of galaxies classed
as field galaxies the field galaxies probably lie close to the
structures traced by the wall galaxies and therefore are not detected
within the void volume. Again, there is no trend seen between the
underdensity and size of voids; the largest voids and the smallest voids all
have consistent values of $\delta \rho / \rho$.

The total volume fraction occupied by voids in the PSCz and UZC surveys is 
30\% and 40\% respectively. Compared to previous estimates, this value is 
low. De Lapparent, Huchra \& Geller (1991) find that high density regions
only fill 25\% of the Center for Astrophysics slices, leaving up to
75\% to be filled by lower density regions. The voids we detect are
extremely low density voids, hence they fill a smaller fraction of 
the survey.

\subsection{Comparison with other papers}

We can compare results with voids detected by EPC97 in the 1.2 Jy Survey. 
The 1.2 Jy survey covers the same area as the PSCz survey but is sparser
due to the higher flux limit.
Consequently, fewer voids are found to the same significance level. 
12 voids with $> 95$\% significance are found and these have an average
diameter of 40$h^{-1}$Mpc. This is significantly larger that the voids
we detect in the PSCz but voids must have a diameter larger than
25$h^{-1}$Mpc to be 95\% significant, thus the smaller voids are not 
reported by EPC97. In most cases, we detect the same regions as 
voids as EPC97. We class void 7 in EPC97 as two separate voids because 
our sample includes fainter galaxies
that restrict the growth of the maximal spheres. There are also two
voids that EPC97 report as significant at the 80\% level that we
do not detect in the PSCz sample (voids 14 and 15 in EPC97).

A recent paper by Plionis \& Basilakos (2001) examines voids in
the PSCz survey. Voids are found by these authors using a smoothed
apparent magnitude limited sample rather than the point
distribution of galaxies and the paper concentrates on measuring the
shapes of voids and making a comparison to various cosmological
models. Plionis \& Basilakos find 14 voids out to 80$h^{-1}$Mpc with
volumes larger than 10$^3 h^{-3}$ Mpc; we find 23 such voids. 9
of the voids are detected using both methods. The remaining 5 voids
lie at distances such that a void with a radius of $ > 10 h^{-1}$Mpc
would not fit fully in the survey geometry. The 14 voids that we
detect but which Plionis \& Basilakos do not detect appear to be 
surrounded by high density regions. The smoothing technique appears to smooth
the galaxies into the voids, restricting the number of voids that can be
detected.

\subsection{No Field Galaxies}
\label{sec:resfield}

What happens if we do not remove field galaxies from the samples?
In either the PSCz or the UZC Survey, we find that 90\% of the same
voids are detected whether or not we make the
wall/field galaxy cut. As expected,  
we miss a couple of the voids when we do not classify galaxies in low
density regions as field galaxies as they interrupt the growth of the
maximal spheres, thus the spheres lie below the detection threshold of 
10$h^{-1}$Mpc and the voids are not detected. 

As the voids we detect when we make the wall/field galaxy cut 
are so empty ($\delta \rho /\rho$ = -0.92, -0.96), 
it is not surprising that there is relatively little difference between
the voids detected with and without applying the wall/field galaxy criteria.

\subsection{Edge Effects}
\label{sec:resedge}

We also compare the volumes of the voids found in our main sample with
voids found in samples that extend 20$h^{-1}$Mpc in depth beyond the
depth of our main samples. We compare the volumes of voids and diameters
of maximal spheres in the center (PSCz) and right hand (UZC) plots in
figure \ref{fig:underest}. This plot shows that in general we
underestimate the volume of voids by a larger factor at larger distance. 
This is not unexpected as it is only voids found at
large depth that should be affected by the survey depth
limit. However, the diameter of the maximal spheres are in reasonable
agreement and the diameters are found to be the same to within 20\%.

\begin{table*}
\begin{centering}
\begin{tabular}{cccccccc}
D(Max-Sphere) & D(Equiv) & Volume & Distance & $\alpha$ & $\delta$ & $\delta \rho / \rho$ & Max-Sphere  \\ 
(h$^{-1}$Mpc) &  (h$^{-1}$Mpc) & (h$^{-3}$Mpc$^3$) &  (h$^{-1}$Mpc) & degrees & degrees &  & Fraction \\ \hline
35.68  & 44.45  &  45992.1  & 56.34  &   355.0  & -47.7  & -0.91  &  0.52  \\
31.34  & 37.12  &  26789.5  & 59.31  &   97.8  & -50.5  & -0.94  &  0.60  \\
31.00  & 39.99  &  33496.0  & 54.44  & 224.9  &  -2.0  & -0.92  &  0.47  \\
30.02  & 31.28  &  16018.4  & 57.57  & 246.9  & -19.6  & -0.96  &  0.88  \\
28.52  & 33.55  &  19773.2  & 59.22  & 242.4  &  33.9  & -0.96  &  0.61  \\
27.29  & 31.87  &  16955.0  & 52.01  & 256.8  &   0.9  & -0.95  &  0.63  \\
27.27  & 32.86  &  18585.3  & 58.63  &  334.3  &   4.8  & -0.94  &  0.57  \\
27.22  & 32.28  &  17606.6  & 56.80  &   61.1  & -27.6  & -0.92  &  0.60  \\
26.88  & 28.85  &  12572.5  & 59.52  &  149.1  &  56.6  & -0.90  &  0.81  \\
26.83  & 36.78  &  26053.0  & 49.03  &  351.9  & -21.1  & -0.94  &  0.39  \\
26.59  & 35.66  &  23749.8  & 41.18  &   34.0  & -55.2  & -0.99  &  0.41  \\
26.41  & 34.71  &  21891.0  & 55.46  &  206.9  &  71.9  & -0.95  &  0.44  \\
26.08  & 32.09  &  17301.4  & 59.26  &  164.1  & -29.3  & -0.94  &  0.54  \\
25.88  & 30.16  &  14357.5  & 58.90  &  209.4  & -42.9  & -0.93  &  0.63  \\
25.73  & 29.27  &  13129.7  & 59.55  &  210.4  &  52.8  & -0.97  &  0.68  \\
24.44  & 31.89  &  16980.8  & 53.96  &  170.3  &  34.5  & -0.92  &  0.45  \\
24.10  & 31.15  &  15821.7  & 43.60  &  310.6  & -29.4  & -0.96  &  0.46  \\
24.03  & 31.27  &  16012.0  & 49.65  &  159.7  &   5.3  & -0.95  &  0.45  \\
23.78  & 30.58  &  14976.6  & 53.82  &  192.8  &  15.9  & -0.91  &  0.47  \\
23.68  & 32.14  &  17384.9  & 52.11  &   49.3  &  15.0  & -0.90  &  0.40  \\
23.53  & 25.72  &   8905.3  & 59.24  &  316.0  & -74.1  & -0.95  &  0.76  \\
23.38  & 26.02  &   9224.0  & 54.02  &   85.1  &  -6.1  & -0.95  &  0.73  \\
23.22  & 26.70  &   9965.7  & 58.41  &   51.5  & -46.0  & -0.91  &  0.66  \\
22.90  & 26.02  &   9223.6  & 59.03  &  285.5  &  47.1  & -0.88  &  0.68  \\
22.82  & 27.82  &  11273.7  & 38.63  &  263.4  &  43.6  & -0.96  &  0.55  \\
22.54  & 24.24  &   7462.0  & 33.63  &  250.3  &  -5.6  & -0.94  &  0.80  \\
22.47  & 27.01  &  10322.9  & 56.80  &  317.2  & -15.7  & -0.96  &  0.57  \\
22.14  & 26.66  &   9923.2  & 58.08  &  124.2  &   6.8  & -0.96  &  0.57  \\
21.09  & 24.57  &   7764.9  & 43.66  &  102.5  &  48.6  & -0.97  &  0.63  \\
20.97  & 20.83  &   4731.8  & 56.08  &  311.6  &  73.4  & -0.86  &  1.02  \\
20.95  & 26.36  &   9594.2  & 56.45  &  223.1  &  31.7  & -0.93  &  0.50  \\
20.66  & 27.51  &  10899.4  & 42.51  &  332.4  &  15.9  & -1.00  &  0.42  \\
20.50  & 23.26  &   6586.8  & 59.68  &   27.5  &  21.4  & -1.00  &  0.69  \\
20.38  & 22.60  &   6047.6  & 46.97  &   91.5  & -22.2  & -1.00  &  0.73  \\
20.21  & 21.26  &   5034.4  & 58.53  &   58.1  & -84.8  & -0.96  &  0.86  \\ \hline
\end{tabular}
\caption{The voids in the PSCz survey, with a distance cut of r $< 80
h^{-1}$Mpc, where field galaxies are differentiated from wall
galaxies. We give (reading left to right) the diameter of the maximal sphere
detected in the void, the equivalent diameter assuming the volume
of the void is spherical, the volume of the void, the location of the
void in distance, $\alpha$ and $\delta$ coordinates assuming an Einstein-de Sitter
Universe, the density contrast of voids and the ratio of the volume of
the largest hole to the total void volume.}
\label{tab:psczvoids}
\end{centering}
\end{table*}

\begin{table*}
\begin{centering}
\begin{tabular}{cccccccc}
D(Max-Sphere) & D(Equiv) & Volume & Distance & $\alpha$ & $\delta$ & $\delta \rho / \rho$ & Max-Sphere  \\ 
(h$^{-1}$Mpc) &  (h$^{-1}$Mpc) & (h$^{-3}$Mpc$^3$) &  (h$^{-1}$Mpc) & degrees & degrees &  & Fraction \\ \hline
29.31  & 39.73  &  32828.0  & 51.22  &  197.8  &  73.9  & -0.95  &  0.40  \\
28.91  & 36.12  &  24667.1  & 42.09  &  247.2  &  39.8  & -0.95  &  0.51  \\
27.31  & 33.79  &  20204.3  & 49.81  &   52.1  &  15.8  & -0.97  &  0.53  \\
27.14  & 32.22  &  17506.9  & 59.56  &  209.7  &  52.6  & -0.97  &  0.60  \\
25.20  & 33.92  &  20443.4  & 44.54  &  168.0  &  38.3  & -0.96  &  0.41  \\
24.83  & 29.25  &  13104.5  & 49.15  &  254.4  &  13.7  & -0.97  &  0.61  \\
23.94  & 28.20  &  11747.8  & 59.68  &  332.5  &  21.8  & -0.96  &  0.61  \\
23.55  & 27.63  &  11045.8  & 47.74  &  196.6  &  12.2  & -0.95  &  0.62  \\
23.17  & 26.75  &  10025.9  & 38.91  &  334.5  &  18.6  & -0.96  &  0.65  \\
23.00  & 29.63  &  13625.9  & 61.75  &  136.9  &  48.4  & -0.97  &  0.47  \\
21.49  & 26.29  &   9509.8  & 45.36  &  162.6  &  14.7  & -0.96  &  0.55  \\
21.18*  & 23.89  &   7142.6  & 61.65  &   3.7  &  42.0  & -0.97  &  0.70  \\
21.07  & 24.15  &   7378.5  & 63.02  &   32.5  &  19.7  & -0.95  &  0.66  \\
20.93  & 25.54  &   8723.1  & 60.93  &  256.2  &  41.7  & -0.97  &  0.55  \\
20.79  & 26.23  &   9454.6  & 51.65  &  277.2  &  57.5  & -0.95  &  0.50  \\
20.55  & 26.78  &  10060.5  & 56.82  &  139.4  &  65.1  & -0.97  &  0.45  \\
20.54  & 23.72  &   6990.5  & 62.58  &  212.2  &  26.4  & -0.97  &  0.65  \\
20.28*  & 21.07  &   4901.1  & 30.35  &   48.6  &  21.9  & -0.97  &  0.89  \\
20.20  & 26.91  &  10203.0  & 51.07  &  141.2  &  25.4  & -0.95  &  0.42  \\ \hline
\end{tabular}
\caption{The voids in the UZC, with a distance cut of r $< 73.6
h^{-1}$Mpc, where field galaxies are differentiated from wall
galaxies. We give (reading left to right) the diameter of the maximal sphere
detected in the void, the equivalent diameter assuming the volume
of the void is spherical, the volume of the void, the location of the
void in distance, $\alpha$ and $\delta$ coordinates assuming an 
Einstein-de Sitter
Universe, the density contrast of voids and the ratio of the volume of the
largest hole to the total void volume. The two voids marked with an asterisk are discussed in the text.}
\label{tab:uzcvoids}
\end{centering}
\end{table*}

\section{Conclusions}
\label{sec:conc}

We have developed and tested a void finding algorithm, similar to that of 
EP97, and applied it to the PSCz and UZC surveys.
Our method differs slightly from that of EP97 in the criteria that we
use to identify unique voids;
there is no obvious definition of how to
group holes into voids. We demonstrate that our technique gives
robust results in the sense that different samples from the same
survey yield the same voids and we detect similar voids from
redshift surveys with different wavelength selection. 
As an extension to the work of EP97 and EPC97 we
quantify the effect of our void definition on the number of
voids detected within a survey and we provide estimates of how
accurately we are able to recover the volume of voids. We determine
that the diameters of the voids presented here are probably accurate to
within 20\%. Detecting voids with relative densities of 
$\delta \rho /\rho =-0.92$,
we find that up to 40\% of the volume in
the surveys under consideration is found in void regions. This is
consistent with the findings of EPC97 and shows that voids are indeed
a large part of the universe.

The next generation of surveys, the 2dF Galaxy Redshift Survey
(Colless et al. 2001 and references therein) 
and, in particular, the Sloan Digital Sky Survey
(York et al. 2000 and references therein) will aid our 
understanding of voids. Both of these
surveys cover a larger sky area than the UZC, although not quite as large
an area as the IR surveys. The SDSS will cover a quarter of the sky in
one contiguous area which will be especially useful for void
detection. Both the 2dFGRS and the SDSS will reach fainter magnitude
limits than previous surveys, approximately 4 magnitudes deeper than
the UZC. This will allow us to construct volume limited samples with
more galaxies, which extend to greater depths. Perhaps more importantly, 
the multiband digital photometry, as well as the
deep spectroscopy,  
of the SDSS will also allow us to
study the properties of the field galaxies in detail. We will have
a large enough sample of galaxies found in low density environments to 
statistically check if they have different properties than the galaxies found
in the wall regions, allowing the role of environment on galaxy 
properties and formation to be examined.

\acknowledgments 
MSV acknowledges support from NSF grant AST-0071201 and the John
Templeton Foundation. We thank Andrew Benson for useful conversations
and the referee for suggesting useful references.

\end{document}